\documentclass{svproc}
\usepackage{url}
\usepackage{multirow}
\usepackage{graphicx}
\usepackage{hyperref}
\usepackage{booktabs, tabularx}
\usepackage{subcaption}
\usepackage[table]{xcolor}  

\usepackage{tikz}
\newlength{\leftimgheight}
\newcommand{\cellsignal}[3][1]{%
	\begin{tikzpicture}[x=#1ex,y=#1ex,baseline={(0,0)}]
		\foreach \i in {1,...,#2} {
			\pgfmathtruncatemacro{\filled}{ifthenelse(\i<=#3,1,0)}
			\pgfmathsetmacro{\h}{\i}                     
			\pgfmathsetmacro{\x}{(\i-1)*1.2}             
			\ifnum\filled=1
			\filldraw[fill=black,draw=black] (\x,0) rectangle ++(1,\h);
			\else
			\draw[fill=gray!25,draw=black] (\x,0) rectangle ++(1,\h);
			\fi
		}
	\end{tikzpicture}%
}

\usepackage{color}

\begin{document}

\mainmatter              
\title{Evaluating 5G-connected IoT for Power Line Temperature Prediction: Real-World Latency and Cost Trade-offs Between MEC and Cloud}
\titlerunning{Latency and Cost Trade-offs between MEC and Cloud}  
\author{Aakash Sharma\inst{1}
	\and
	Sigmund Akselsen\inst{2}
	\and
	Anders Andersen\inst{1}
	\and
	Lars Ailo Bongo\inst{1}
	\and
	Arne Munch-Ellingsen\inst{2}
} 
\authorrunning{Aakash Sharma et al.} 

\institute{UiT -- The Arctic University of Norway, Tromsø, Norway,\\
\email{aakash.sharma@uit.no}
\and
Telenor Research, Fornebu, Norway}

\maketitle              
\begin{abstract}
One of the key promises of Mobile Edge Computing~(MEC) is its low latency. 
Current large-scale IoT deployments rely on cloud for their reliability, low cost, and ease of use. 
For outdoor IoT deployments, 5G cellular networks offer significantly enhanced bandwidth and dramatically reduced latency compared to previous generations, enabling real-time data processing and control.  
Therefore, leveraging 5G connectivity is crucial for outdoor IoT applications requiring responsiveness and complex data handling. 
Combining MEC with 5G has the potential to provide the ease of cloud computing alongside low latency.  
We investigate the latency performance on a 5G cellular network with an experimental MEC setup. 
In our proof-of-concept, we demonstrate the benefits of using an edge-based compute server for real-time power transmission line analytics. 
We compare our solution with state-of-the-art multi-region cloud deployments and discuss the advantages of mobile edge computing~(MEC). 
Our real-world evaluation demonstrates a low latency of 44.62 ms for MEC compared to cloud regions; however, the gap is narrowing. 
While such low latencies can benefit real-world deployments, they remain insufficient to meet the stringent requirements of smart power grid operations ($\sim$8 ms).

\end{abstract}

\section{Introduction}
Mobile Edge Computing, or Multi-Access Edge Computing~(MEC), introduces a novel architecture that enables computing closer to the edge of the network, reducing latency and enabling real-time processing capabilities. 
A key feature of MEC is its ability to facilitate low-latency communication between endpoints, making it attractive for applications requiring low latencies. 
The fifth generation~(5G) cellular networks provide a complementary infrastructure, offering faster network speeds, lower latency, and increased bandwidth to support the growing number of devices connecting to the Internet. 
The Internet of Things~(IoT) is a notable beneficiary of 5G, enabling seamless integration of a vast array of devices into our digital landscape. 
Furthermore, recent advancements in machine learning and artificial intelligence have the potential to augment the intelligence and control of IoT devices, unlocking new possibilities. 
The widespread availability of 5G networks provides a unique opportunity to test and validate the benefits of MEC in real-world scenarios, which is crucial for its large-scale adoption. 

Multi-access Edge Computing~(MEC) targets a new class of applications and services with stringent requirements on ultra-low latency, high bandwidth, and efficient resource utilization~\cite{filali2020multi}. 
Although multiple architectures and approaches have been proposed to reduce latency in MEC, real-world evaluations are required to determine which classes of low-latency applications can be supported by current 5G networks~\cite{filali2020multi,shahzadi2017multi}. 
Recent studies indicate that 5G and MEC architectures are being designed to support distributed AI/ML workloads by enabling low-latency, real-time processing close to data sources~\cite{kaloxylos2021ai}. 
The increasing complexity of next-generation networks and services necessitates AI/ML-driven automation and optimization, making MEC a key enabler for future intelligent applications. 

This study is set in Norway, a unique test environment due to the widespread availability of 5G, varied landscape, and remote communities in Northern Norway. 
We focus on IoT devices that connect directly to 5G networks, excluding those that rely on home Wi-Fi connections. 
This study investigates the differences in P99 latency and operational costs between local edge and regional cloud deployment architectures for mobile edge applications operating on Telenor's 5G network in Norway.

We identify a promising application of power transmission line sensors, which gather data on power transmission lines and cable condition. 
This data is utilized to predict Ambient Adjusted Ratings~(AAR), enabling more efficient use of existing power transmission lines. 
A study in Argentina~\cite{pascual2020potential} demonstrates the potential of AAR to increase transmission capacity on thermally limited lines, thereby unlocking the full output of a wind-based electricity plant. 

The contributions of this paper are:
(a) real-world measurements of P99 latency~\cite{dean2013tail} for an end-to-end proof-of-concept~(PoC) machine learning application deployed across multiple regional cloud and local edge~(MEC) environments on Telenor's 5G network in Norway,
(b) a comparative analysis of the operational costs associated with these deployment architectures, and
(c) an evaluation of the scalability of the PoC  machine learning application, analyzing the relationship between latency and request rate (requests per second).

\section{Related Works}\label{related}
Miladinovic et al.~\cite{miladinovic2021multi} showed that Mobile Edge Computing~(MEC) can significantly reduce latency and jitter compared to cloud deployments, highlighting physical distance and network hops as key latency contributors. 
Huang et al.~\cite{huang2018edge} conducted a comparison of real-time smart grid monitoring on edge and cloud platforms. 
They demonstrated a reduction in detection delay by a factor of 10 by using edge computing versus a cloud platform. 
Deng et al.~\cite{deng2022software} evaluated distributed MEC scenarios through simulation and reported end-to-end latencies of approximately 900 ms for complex image processing tasks, underscoring the challenges of latency-sensitive workloads. 
EdgeBench~\cite{das2018edgebench} measured end-to-end latency for representative edge applications on resource-constrained devices, but it does not capture the capabilities of a modern 5G-connected MEC platform.

Hybrid MEC-cloud architectures have been proposed to improve resilience and performance~\cite{dreibholz2019mobile}.
However, the availability of public 5G networks and MEC by telecom operators changes how such systems should be evaluated. 
In contrast to prior works, we measure real-world end-to-end network latency using a proof-of-concept machine learning inference application to assess which low-latency services can benefit from MEC.

\section{Mobile Edge Computing for Critical Infrastructure}\label{resilience}
Mobile Edge Computing~(MEC) brings compute and storage resources closer to where data is generated, enabling localized processing at the network edge. 
By reducing reliance on distant centralized cloud infrastructures, MEC can significantly lower end-to-end latency and improve service availability. 
Beyond performance benefits, MEC also offers resilience advantages by allowing critical services to continue operating even during partial network failures or connectivity disruptions.

Huang et al.~\cite{huang2018edge} show that smart grid monitoring requires fast, reliable processing of large volumes of sensor data, which centralized cloud solutions struggle to support due to latency and bandwidth constraints. 
Their work demonstrates that placing computation at the network edge enables real-time fault detection, rapid response, and improved system stability by reducing end-to-end delays and dependency on wide-area connectivity. 
This makes edge computing particularly suitable for critical infrastructure, where timely decisions and operational resilience are essential.

Although many existing systems do not yet require ultra-low latency, the increasing digitization of critical infrastructure introduces stricter requirements on reliability, responsiveness, and autonomy. 
In this context, MEC provides a cost-effective alternative to large-scale cloud deployments by enabling localized processing while reducing dependency on long-haul communication links~\cite{zhao2019edge}. 

\section{Use Case: Smart Grids}\label{smartgrids}
Traditional power transmission lines operate under static current limits to prevent thermal overload. 
Rising electrification and energy demand have stressed existing grids, often requiring costly upgrades. 
Recent advances in IoT sensors and power electronics show that existing lines can safely carry more current than static limits allow under favorable conditions~\cite{pascual2020potential}. 
In northern Norway, cold Arctic winters improve heat dissipation, enabling higher transmission capacity, while high summer temperatures in regions like Texas can increase line sag and wildfire risk. 
Some sensors include cameras for early fire detection, allowing rapid intervention.

Safe transmission depends on multiple factors. While grid voltages and topology are predetermined, current limits are traditionally static. 
Modern approaches such as Ambient Adjusted Rating~(AAR) and Dynamic Line Rating~(DLR) dynamically calculate safe limits using real-time measurements and environmental data, potentially increasing line capacity by up to 50\%~\cite{pascual2020potential}.

Our industry partner deploys sensors that transmit real-time data via Telenor cellular network to cloud analytics pipelines using MQTT~\cite{mqtt}. 
These systems combine historical data, weather forecasts, and live measurements to predict next-day safe AAR/DLR values. 
While grid control remains manual, these insights support day-ahead planning and monitoring.

Future intelligent grids will require widespread sensor deployment for continuous monitoring and automated control. 
Low-latency communication and processing will be essential for rapid responses to changing conditions, including temperature fluctuations, line sag, and fire hazards. 
Real-time, intelligent monitoring will be crucial to safely maximize transmission efficiency, reduce infrastructure costs, and enhance operational resilience in next-generation power grids.

\begin{figure}[!bp]
\centering
\includegraphics[width=.8\textwidth]{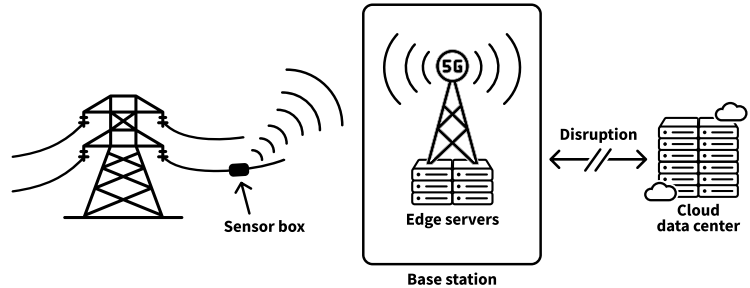}
\caption{Experimental setup showing a sensor box connected via 5G to edge servers and cloud back-end.}
\label{fig:5gedge}
\end{figure}

\subsection{Sensors Data and Machine Learning}\label{sensors}
Cable-mounted sensor boxes (\autoref{fig:5gedge}) enable continuous monitoring and data collection for advanced analytics on power transmission lines. 
They typically measure conductor temperature and current, wind speed and direction, and line sag. 
Line sag can be affected by snow and ice in winter, and these measurements aid maintenance planning. 
Some sensor boxes also include cameras to detect anomalies and hazards, such as forest fires, enabling timely preventive actions. 

We received a year of sensor readings to train our model, with sensitive information such as GPS coordinates removed. 
The dataset contains ambient temperature, line temperature (both standard and thermistor-measured), line current, atmospheric pressure, humidity, and distance to the ground. 

Using this dataset of 947,590 records, we trained a feedforward neural network to predict linetemp and linetemp\_ntc from multivariate sensor data. 
The model has an input layer with eight features (including temporal features extracted from the timestamp), three hidden layers with 256, 128, and 64 neurons, and an output layer predicting the two target variables. 
The data were partitioned into training~(70\%), validation~(10\%), and test~(20\%) sets, with all features standardized using z-score normalization.

\subsection{Proof-of-Concept: Temperature Prediction}\label{edge-compute}
In this proof-of-concept, we simulate a power transmission line sensor that periodically sends a tuple of parameters to a server, which in turn infers the temperature of the power transmission line. 
Rather than computing full AAR/DLR values—which require large streams of historical and contextual data—we design a lightweight inference task that reflects a common IoT interaction pattern: multiple devices transmitting small data tuples and receiving near-real-time responses. 
This setup also represents intelligent IoT devices that rely on server-side inference to guide subsequent actions, with device state optionally maintained on the server. 
Our primary objective is to analyze end-to-end latency to assess the feasibility of low-latency IoT applications.

After training the machine learning model, we implemented a containerized web server that receives sensor data, performs inference, and returns predicted line temperature values (\textit{linetemp} and \textit{linetemp\_ntc}). 
The server was implemented using FastAPI with Uvicorn and packaged as a Docker image. 

The system architecture consists of two main components: (i) periodic network latency measurements, performed hourly, and (ii) inference-based temperature prediction using the trained model. 
Our evaluation aims to assess the feasibility of low-latency to support interactive monitoring and control.
Client-side test scripts are configured via a configuration file specifying server endpoints at different deployment locations. 
The client randomly selects tuples from the sensor dataset (\autoref{sensors}) and sends them to the server, which performs inference and returns the results. 
The client measures round-trip time~(RTT), while the server records inference execution time. 
Although prediction error is computed, it is not analyzed in this paper, as our focus is on network and deployment-induced latency characteristics rather than model accuracy.

RTT and server processing times are measured in milliseconds. 
To ensure comparability, the experimental setup is kept consistent across all evaluations (see \autoref{evaluation}). 
In some cases, the connection falls back to 4G when 5G coverage is unavailable.
This behavior reflects real-world deployments where 5G coverage is primarily available in densely populated areas.

\subsection{Equipment}\label{equipment}
\begin{figure}[htb]
    \centering
    \settoheight{\leftimgheight}{%
        \includegraphics[width=.635\linewidth]{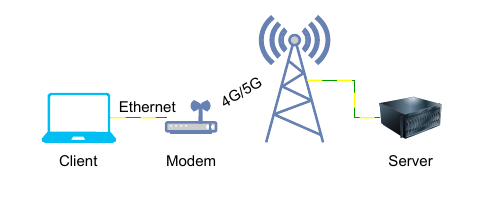}%
    }

    \begin{tabular}{@{}c c@{}}
        \includegraphics[width=.635\linewidth]{images/setup_diagram.pdf}
        &
        \includegraphics[height=\leftimgheight]{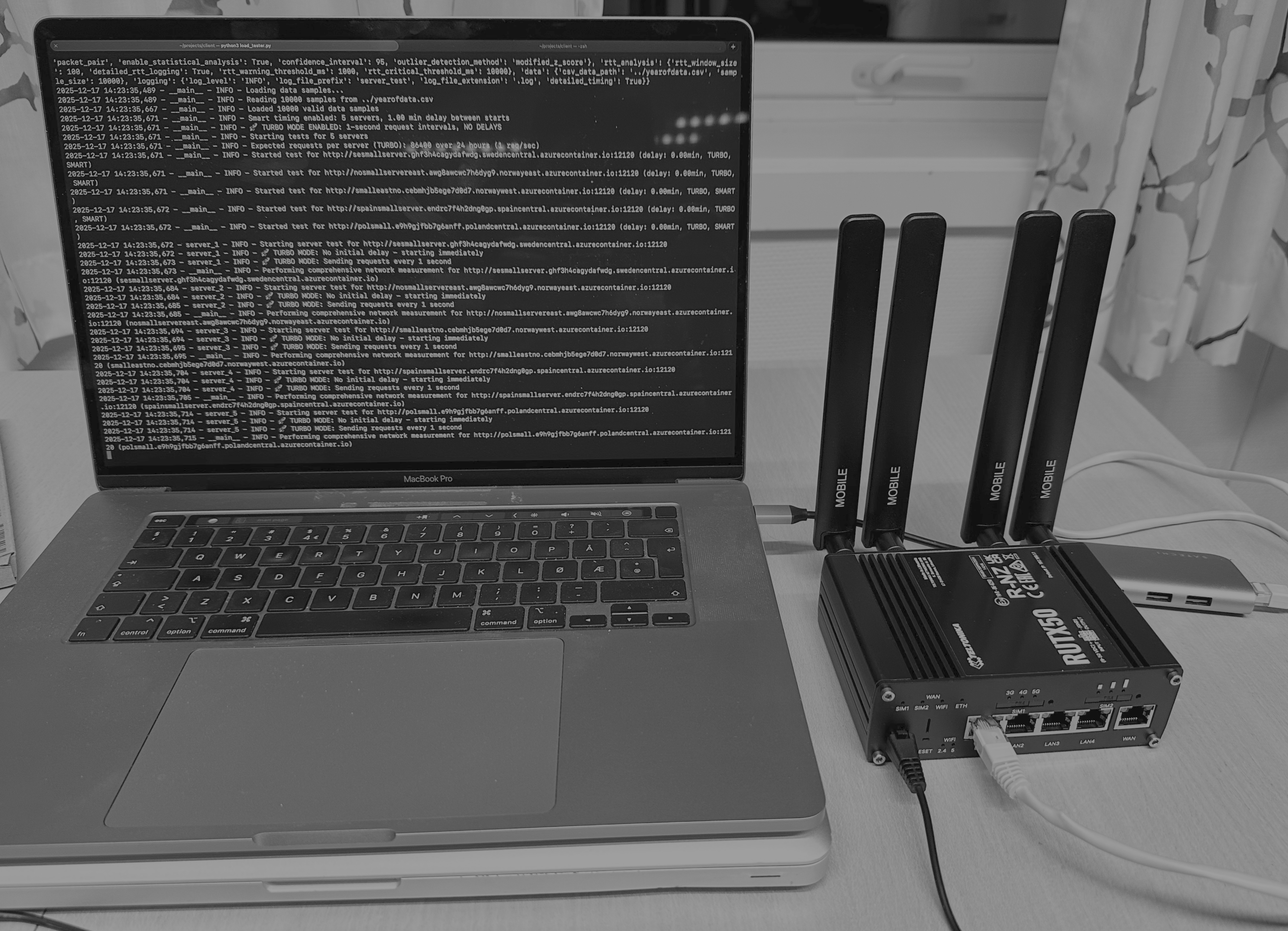}
    \end{tabular}

    \caption{Setup used for latency measurements.}
    \label{fig:setup}
\end{figure}

\autoref{fig:setup} shows our setup, simulating an IoT device sending a JSON payload via POST requests to the server. 
We simulated IoT clients using a MacBook (model A2141) running macOS 26 with Python 3.12, connected to a Teltonika RUTX50 modem over an Ethernet cable. 
LAN port 1 was used for all experiments over 5G, while the WAN port connected to an optical fiber link when testing fiber-based communication. The router firmware was 07.18.3, and the modem version was RG501QEU-AAR12A11M4G with kernel 6.6.96.

\section{Evaluation}\label{evaluation}
We replicate a processing pipeline to investigate the flow of sensor readings as a JSON object from the client to the server and inferred values back to the client. 
We containerized the workflows, enabling easy deployment across cloud and edge~(MEC) environments.

For our evaluation, we focus on two key parameters: latency and cost. 
Latency is critical for time-sensitive applications, such as AR/VR gaming and intelligent robots ($\sim$100 ms)~\cite{zhou2019edge}, industrial automation ($\sim$50 ms)~\cite{stefanovic2018industry}, and power grid systems ($\sim$8 ms)~\cite{rudy2018iec}. 
Cost is a key factor in scaling deployments for large distributed sensor networks. 
We also evaluate latency degradation under increasing request rates.

We evaluated a full day of compute requirements for the analytical pipeline with back-ends deployed in different locations and cloud zones. 
Norway's elongated geography results in large physical distances between remote communities and available cloud regions.
Among major cloud providers, Microsoft Azure was selected as it offers the required regional coverage, including two regions in Norway, namely Norway East~(NO-East) and Norway West~(NO-West).
NO-East is located near Oslo, while NO-West is near Stavanger.
Both data centers are more than one thousand kilometers from Tromsø. 
We also included Sweden Central~(SE-Central) cloud region, reflecting the current compute pipeline used by the sensor box developer. 

\subsection{Client Locations}\label{loc}
The locations were selected to simulate different sensor proximities to 5G towers: two in dense urban areas with multiple towers, and one near a defunct hydroelectric power station, representing a remote scenario. 
Building a functional 4G/5G mobile network test environment is challenging due to its inherent complexity~\cite{dreibholz2020flexible}. 
Therefore, for MEC experiments, we relied on Telenor's facilities in Fornebu.
Tests were performed using a public 5G network, and \autoref{tab:network} summarizes the locations, signal strengths, and cellular parameters. 
For comparison, we also tested from a recently established fiber connection, connecting the computer via the router to replicate router-induced latencies.

\begin{table}[t]
\centering
\caption{Cellular network details across locations, as reported by the router.}
\label{tab:network}
\begin{tabularx}{\textwidth}{
  >{\raggedright\arraybackslash}X
  >{\centering\arraybackslash}X
  >{\centering\arraybackslash}X
  >{\centering\arraybackslash}X
  >{\centering\arraybackslash}X
  >{\centering\arraybackslash}X
  >{\raggedright\arraybackslash}X  
}
\toprule
\textbf{Location} (Postal code, City) & 
\textbf{Status Lights} & 
\textbf{RSSI} (dBm) & 
\textbf{RSRQ} (dB) &
\textbf{Network Type} &
\textbf{Connected Bands} &
\textbf{Carrier Aggregation }\\
\midrule
9019, Tromsø & 
5G {\color{green!70!black}\textbullet}  4G {\color{green!70!black}\textbullet}  &
-60 & 
-8 & 
5G (NSA) & 
LTE B1/B3/B7, 5G n78 &
Active  \\
0366, Oslo & 
5G {\color{green!70!black}\textbullet}  4G {\color{green!70!black}\textbullet}  &
-45 & 
-10 &
5G (NSA) & 
LTE B1/B3/B7, 5G n78 &
Active  \\
9131, Kårvik & 
5G {\color{green!70!black}\textbullet}  4G {\color{green!70!black}\textbullet}  &
-76 & 
-20 &
4G (LTE) & 
LTE B20&
Inactive  \\

1360, Fornebu & 
5G {\color{green!70!black}\textbullet}  4G {\color{white!70!black}\textbullet}  & 
-47  & 
-10  &
5G (SA) & 
5G n78 &
Inactive \\
\bottomrule
\end{tabularx}
\end{table}
    
\section{Results}\label{results}
We conducted experiments in which the client machine sent requests and received responses from servers hosted in various cloud regions and at the edge. 
RTT data were collected and analyzed to plot key performance indicators. 
Statistical significance was validated using \textit{p}-value tests, and the coefficient of volatility

\begin{equation}
CV = \frac{\sigma}{\mu}
\end{equation}
was calculated to quantify latency variability. 
All scripts and observed latency logs are publicly available in Zenodo. 
A summary of the evaluation is also presented in \cite{sharma_2026_18435283}.

\subsection{Latency}\label{latency}
We ran our program to collect round-trip time~(RTT) measurements over a period of 24 hours from each location. 
We analyzed the results and highlighted differences up to and beyond the 99th percentile by plotting P99 latencies. 
P99 is often used as a key performance metric for measuring latencies for latency-sensitive applications~\cite{hoiland2016measuring}. 
Also, a high percentile like P99 is critical for assessing the network's unpredictability~\cite{iorio2021latency}. 

\begin{figure}[!htb]
  \centering
  \begin{subfigure}{\linewidth}
    \centering
    \includegraphics[width=\linewidth,height=0.28\textheight,keepaspectratio]{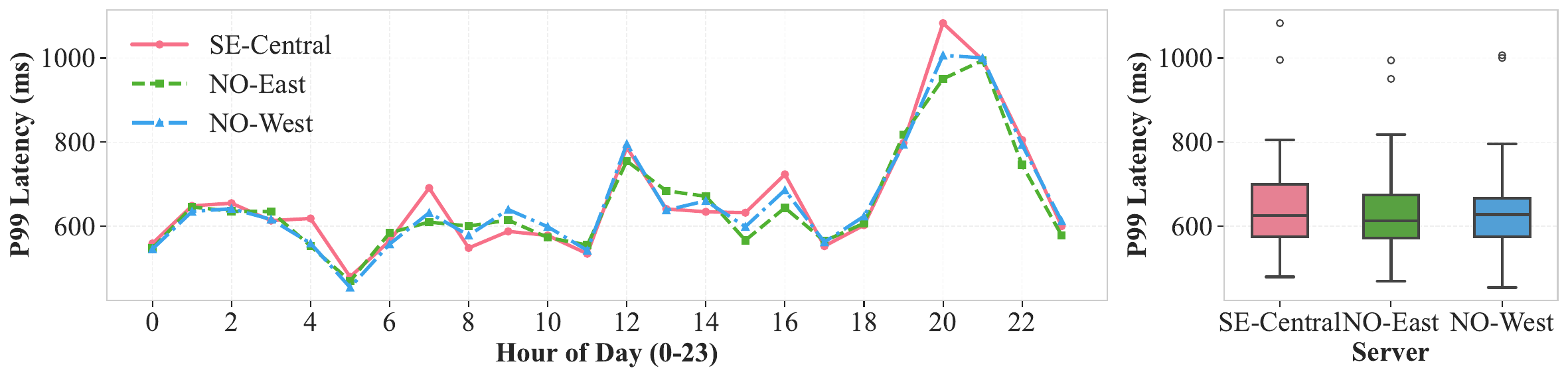}
    \caption{9131, Kårvik}
    \label{fig:a}
  \end{subfigure}\par
  \begin{subfigure}{\linewidth}
    \centering
    \includegraphics[width=\linewidth,height=0.28\textheight,keepaspectratio]{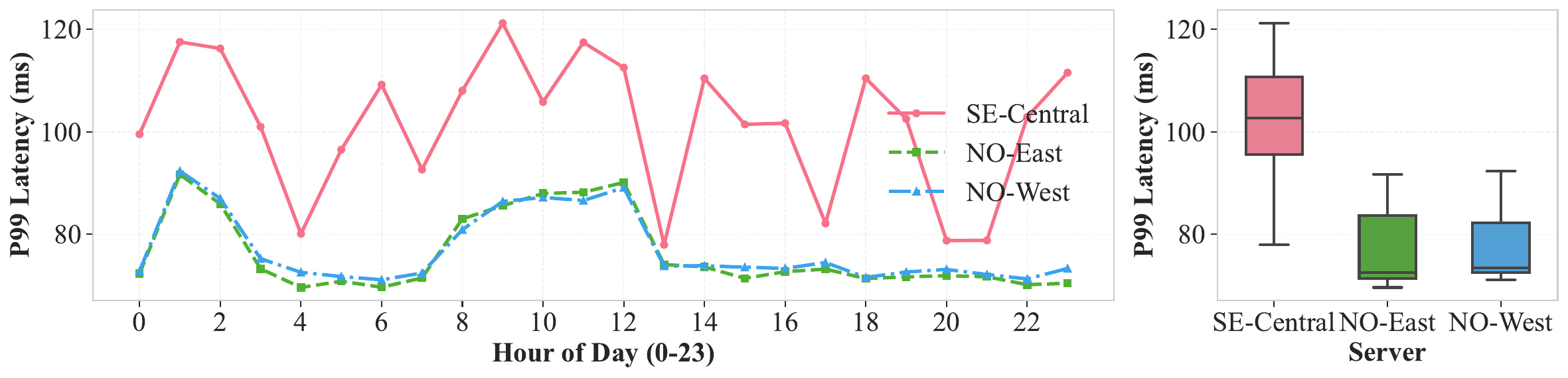}
    \caption{9019, Tromsø}
    \label{fig:b}
  \end{subfigure}\par
  \begin{subfigure}{\linewidth}
    \centering
    \includegraphics[width=\linewidth,height=0.28\textheight,keepaspectratio]{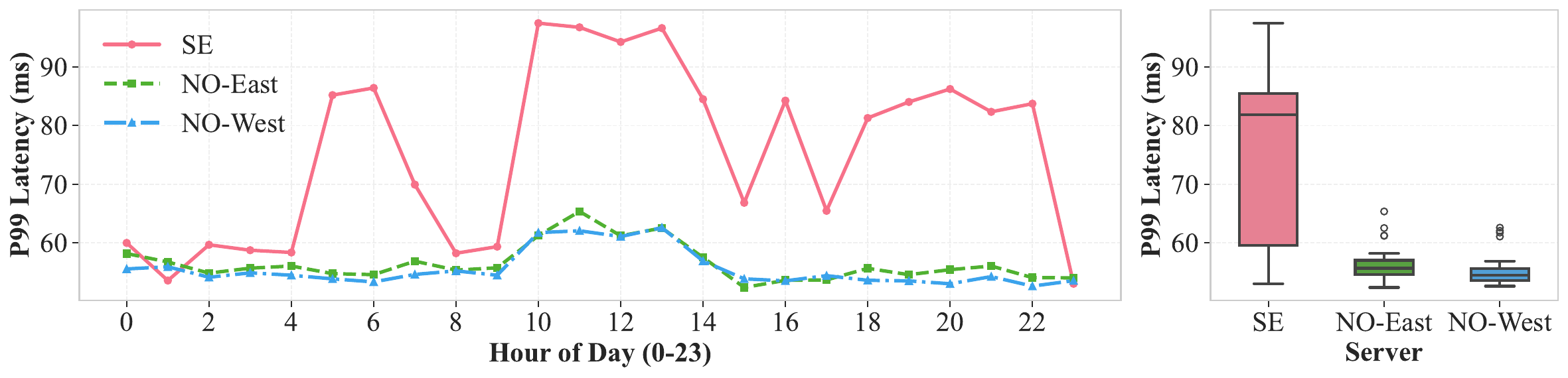}
    \caption{0366, Oslo}
    \label{fig:c}
  \end{subfigure}

  \caption{P99 latency from multiple Norwegian locations to cloud regions over Telenor's cellular 4G/5G network.}
  \label{fig:overall}
\end{figure}

\autoref{fig:a} shows the end-to-end latency from Kårvik. 
The coverage is spotty in this area and resembles locations near the edge of cellular service coverage. 
\autoref{tab:network} shows the limited cell signal strength and the lack of 5G. 
There is not much difference between the latencies to different cloud regions, as signal propagation time is likely the main contributing factor to the high latencies observed. 
There is also high relative volatility ($\approx$0.2) for each cloud region. 
The latency for each cloud region is around 660 ms, which is high for latency-sensitive applications. 
We performed a \textit{p}-test and obtained a \textit{p}-value of 0.9726.
Therefore, there is no significant difference between servers since \textit{p} $\geq$ 0.05. 

From UiT campus in Tromsø, we observed latency values plotted in \autoref{fig:b}. 
The signal strength is good. 
We can clearly see the difference between cloud servers in Norway~(NO-East and NO-West) and Sweden~(SE-Central). 
On average, the difference in latency between the Norwegian regions and Sweden was around 30 ms. 
The volatility was higher for SE-Central~(0.132) compared to NO-East~(0.099) and NO-West~(0.089). 
We obtain a \textit{p}-value of 7.1199e-09; therefore, there are significant differences between servers (\textit{p}$ < 0.05$). 

From the client machine in Oslo, the latency was further reduced to each of the cloud regions. 
There was higher volatility (0.201) in the latency to SE-Central, but it was very stable for the other two regions (NO-East and NO-West). 
Average end-to-end latencies to the NO-East and NO-West regions were $\approx$58 ms. 
The average latency was $\approx$85 ms for the SE-Central region. 
We observed a \textit{p}-value of 3.6124e-07 and therefore the difference between servers is significant (\textit{p}$ < 0.05$).

\begin{figure}[tb]
    \centering
    \includegraphics[width=\linewidth]{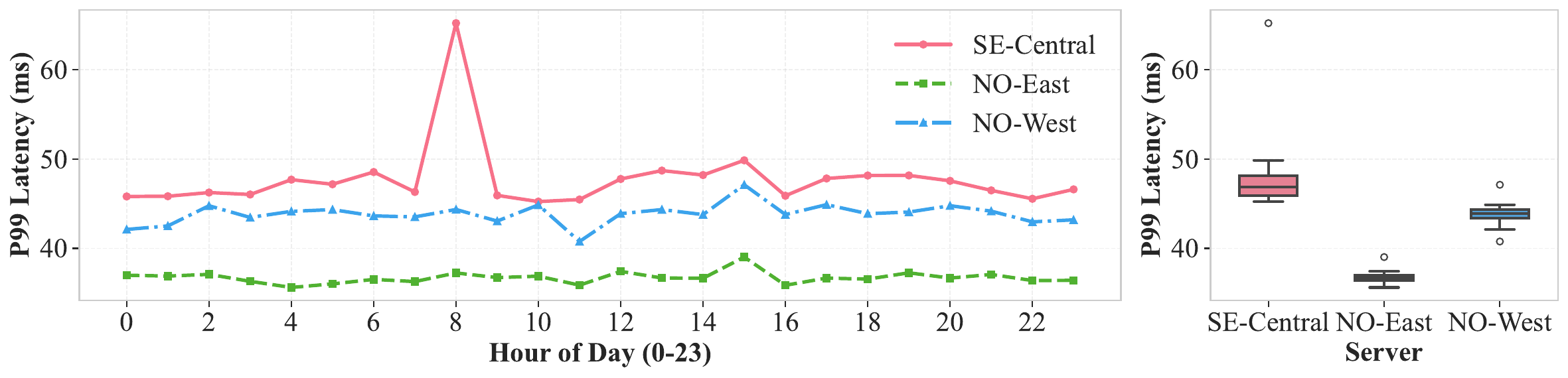}
    \caption{P99 latency from the client to different cloud regions over a fiber-optic network.}
    \label{fig:fiber}
\end{figure}

Additionally, we conducted the same evaluation using a fiber-optic-based connection. The latency results are shown in \autoref{fig:fiber}. 
As expected, the Norwegian regions exhibited lower latency than the Swedish region; however, the difference in average latency was less than 10 ms. 
Despite this small absolute difference, the measured \textit{p}-value of 3.7374e-14 indicates a statistically significant difference between the latencies.

\begin{figure}[htb]
    \centering
    \includegraphics[width=\linewidth]{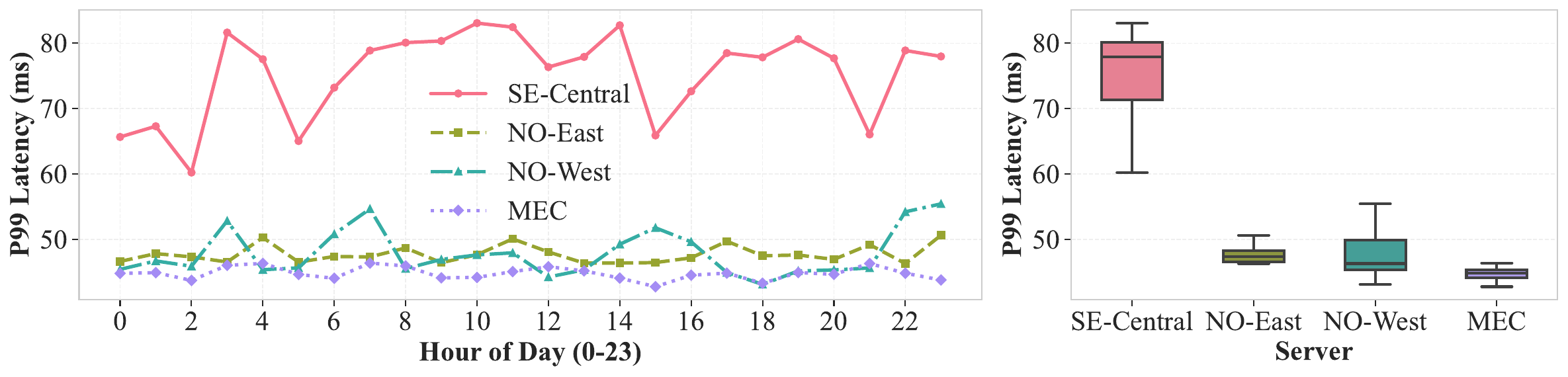}
    \caption{P99 latency comparison between cloud regions and the edge in Telenor's experimental 5G MEC network.}
    \label{fig:mec}
\end{figure}

Finally, we measured the latencies to all the cloud regions and to a server deployed at the mobile edge~(MEC) close to the 5G core. 
\autoref{fig:mec} shows the latencies observed. 
MEC shows the lowest average latency at approximately 44.62 ms, with a volatility of 0.021. 
The average latency to NO-East is 47.69 ms, followed by 48.80 ms for NO-West and 78.49 ms for SE-Central. 
Compared to MEC, latency is 7\% higher for NO-East, 9\% higher for NO-West, and 76\% higher for SE-Central.
Similar to the measurements in Tromsø, the difference between SE-Central and NO-East is $\approx$30 ms. 
Again, the volatility is higher for SE-Central. 
We measured a \textit{p}-value of 1.2542e-15, indicating significant statistical differences in latency among servers.

\subsection{Capacity Planning}\label{capacity}
\begin{figure}[htb]
    \centering
    \includegraphics[width=.95\linewidth]{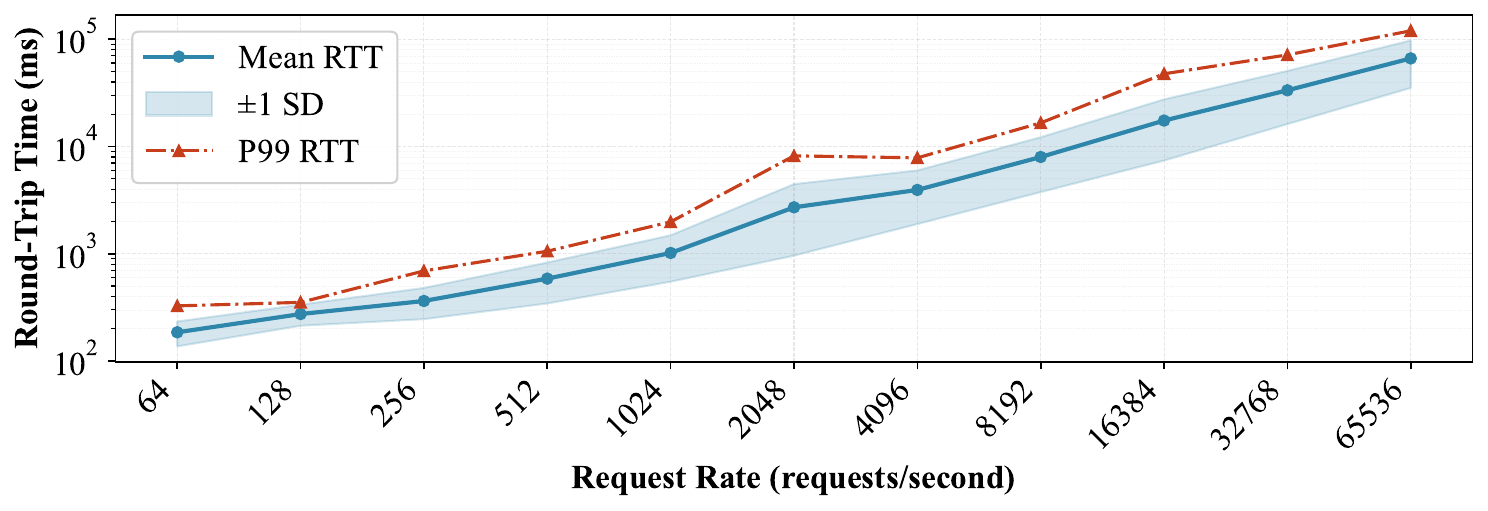}
    \caption{Latency degradation under increased requests.}
    \label{fig:capacity}
\end{figure}

In our P99 evaluation, we initially sent approximately one request per second to the server. 
We assessed how end-to-end latency degrades under increasing load by gradually increasing the request rate. 
The experiment was conducted from Tromsø to the SE-Central region, and the results are shown in \autoref{fig:capacity}. 
At 1024 requests/second, the average latency reached approximately 1 second. 
At 8,192 requests/second, some requests began to fail, and the fraction of failed requests increased further as the load continued to rise. 
At 65,536 requests/second, the P99 latency reached 119 seconds, far exceeding the requirements of interactive applications. 
The observed latency degradation is useful for planning server scaling to support higher request rates.

\subsection{Cost}\label{cost}
We used Azure cloud services to host containerized workloads across multiple regions. 
Azure employs dynamic pricing, and the Norwegian krone~(NOK) is subject to volatility. 
Each container was configured with minimal resources (1 CPU and 512\,MB memory). 
CPU utilization remained within limits, except during capacity testing (\autoref{capacity}), where some requests failed. 

The average daily cost of running a container was 10.4~NOK in SE-Central, 12.09~NOK in NO-East, and 15.71~NOK in NO-West. 
In addition, a container registry was required for orchestration, incurring a fixed cost of 1.36~NOK per day independent of usage. 
Running a container continuously for one year therefore costs 4,292~NOK in SE-Central, 4,909~NOK in NO-East, and 6,231~NOK in NO-West, corresponding to increases of 14\% and 45\% relative to SE-Central. 
Currently, no pricing information is available for Telenor's MEC offering.

\section{Discussion}\label{discussion}
Our results indicate that MEC over 5G can achieve latencies comparable to those of optical fiber connections, enabling low-latency applications even in remote areas with cellular coverage. 
MEC shows lower latency volatility than cloud services, resulting in a more consistent and improved user experience~\cite{dean2013tail}.  
Hybrid MEC-cloud architectures provide support for intelligent applications with low latency while maintaining the scalability benefits of cloud platforms~\cite{dreibholz2019mobile,taherkordi2018future}. 
Furthermore, MEC facilitates AI/ML deployments by allowing low-latency execution and dynamic resource management~\cite{kaloxylos2021ai}.

For critical infrastructures, such as power grids, ultra-low latency is crucial for automated monitoring and control~\cite{rudy2018iec}. 
MEC, in conjunction with cloud resources, has the potential to enhance resilience in these critical systems. 
This hybrid approach can improve reliability and responsiveness in scenarios where rapid decision-making is essential.

\section{Conclusion}
In this paper, we analyzed real-world latency and cost of deploying an analytics pipeline in both multi-region cloud and edge environments over 5G. 
The results demonstrate that mobile edge computing~(MEC) can significantly reduce end-to-end latency (44.62 ms) compared to multi-region cloud deployments, benefiting latency-sensitive IoT applications such as smart power grids. 
However, present-day 5G MEC solutions remain insufficient for meeting the ultra-low latency requirements ($\sim$8 ms) of smart grid operations. 
Future progress in edge architectures and network capabilities will be essential to enable reliable, real-time grid intelligence.

\subsubsection{\ackname}
This work was supported by the AI4Europe project under EU Horizon Europe Grant No. 101070000. The authors thank Abdelhakim Cherifi for his support.

\bibliographystyle{spmpsci}
\bibliography{references}

\end{document}